\documentclass[12pt,a4paper]{article}

\usepackage[intlimits]{amsmath}
\usepackage{amssymb}
\usepackage{graphicx}
\usepackage[dvips,a4paper,scale={.75,.8}]{geometry}
\usepackage{bbm}
\usepackage{mathrsfs}
\usepackage{axodraw}
\usepackage{wrapfig}
\usepackage{caption}
\usepackage{subfigure}
\usepackage{enumerate}
\usepackage{makeidx}

\usepackage{cancel}

\def\s1{\hat{s}_1}
\def\T1{\hat{t}_1}
\def\t2{\hat{t}_2}
\def\U1{\hat{u}_1}
\def\u2{\hat{u}_2}

\newcommand{\comment}[1]{}

\usepackage{epsfig}

\usepackage{amssymb}

\begin{document}



\begin{center}
{\Large \bf Higgs-radion similarity in production processes involving off-shell fermions
\hfill\\}
\end{center}
\vspace{0.5cm}


\begin{center}
{E.~Boos$^{1}$, S.~Keizerov$^{1}$, E.~Rahmetov$^1$, K.~Svirina$^{1,2}$\\
\hfill\\
{\small \it $^1$Skobeltsyn Institute of Nuclear Physics, Lomonosov Moscow State University}\\
{\small \it  Leninskie Gory, 119991, Moscow, Russia} \\
 {\small \it $^2$Faculty of Physics, Lomonosov Moscow State University,
 Leninskie Gory}\\
{\small \it 119991, Moscow, Russia}  }
\end{center}





\begin{abstract}
The appearance of the radion field, which is associated with the
spin-0 metric fluctuations combined in some manner with the scalar
stabilizing field, and of the radion, the corresponding lowest
Kaluza-Klein (KK) mode, is a generic prediction of stabilized
brane world models. In such models the radion plays the role of
the dilaton, and its mass may be somewhat smaller than that of all
the KK modes of other particles propagating in the
multidimensional bulk. Due to its origin, the radion couples to
the trace of the energy-momentum tensor of the Standard Model, the
interaction Lagrangian of the radion and the Standard Model
fermions being similar to that of the SM Higgs-fermion
interactions except for additional terms, which come into play
only in  the case of off-shell fermions. In the present paper it
is shown that all the contributions to perturbative amplitudes of
physical processes due to these additional terms are canceled out
for both massless and massive off-shell fermions. Thus the
additional fermion-radion terms in the interaction Lagrangian do
not alter any production and decay properties of the radion
compared to those of the Higgs boson.
\end{abstract}

\comment{
}



\section{Introduction}

There are few generic predictions in brane world models as
possible extensions of the Standard Model (SM). The fields
propagating in the multidimensional bulk  manifest themselves  as
KK towers of states on the brane, where we are supposed to live.
These KK tower of states are an example of such generic
predictions. Another phenomenon of this kind is the presence of a
new scalar field (or fields) on "our" brane (TeV-brane) associated
with spin-0 fluctuations of the metric component corresponding to
extra space dimension. The size of the extra space dimension
should be stabilized in order to get a physically meaningful
picture. This leads to the violation of the dilatation invariance,
and the spin-0 component of metric fluctuations being combined in
some manner with the additional scalar field introduced for
stabilizing the size renders a new scalar field called the radion
field. This radion field plays the role of the dilaton in the
system at hand.

The Randall-Sundrum setup \cite{Randall:1999ee} with stabilizing
scalar field as proposed by Goldberger and Wise
\cite{Goldberger:1999uk} or worked out by DeWolfe, Freedman,
Gubser and Karch \cite{DeWolfe:1999cp}, taking into account the
backreaction by solving exactly the coupled equations for the
metric and the scalar field, provides an example of a concrete
realization of a stabilized brane world model. In a number of
studies it was argued that the lowest scalar state, usually called
the radion \cite{Goldberger:1999un, Csaki:1999mp,
Charmousis:1999rg}, might be significantly lighter than all the
other KK excitations \cite{Csaki:2000zn, Boos:2004uc}.

The phenomenology of the radion follows from the simple effective
Lagrangian
\begin{equation}
 L = - \frac{r(x)}{\Lambda_r} T_{\mu}^{\mu},
 \label{L1}
\end{equation}
 where $T_{\mu}^{\mu}$ is the trace of the SM energy-momentum tensor,
 $r(x)$ stands for the radion field, and
${\Lambda_r}$ \footnote{For the parameter $\Lambda_r$, the
notation $\Lambda_{\phi}$ is also often used. $\Lambda_r$ and
another often used parameter $\Lambda_{\pi}$ are related by
$\Lambda_r = 2\sqrt{3} \Lambda_{\pi}$.} is a dimensional scale
parameter. Since the radion plays the role of the dilaton, it
interacts with the massless photon and gluon fields via the
well-known dilatation anomaly in the trace of the energy momentum
tensor. At the lowest order in the SM couplings the trace has the
following form, where the fields, as supposed in most of the
studies,  are taken on the mass-shell,
\begin{equation}
T_{\mu}^{\mu} =
\frac{\beta(g_s)}{2g_s}G^{ab}_{\rho\sigma}G_{ab}^{\rho\sigma} +
\frac{\beta(e)}{2e}F_{\rho\sigma}F^{\rho\sigma} -
m^{2}_{Z}Z^{\mu}Z_{\mu} - 2 m^{2}_{W}W^{+}_{\mu}{W^{-}}^{\mu}  +
\sum_f m_f \bar{f} f,
 \label{Tr1}
\end{equation}
and the sum is taken over all SM fermions, $\beta(g_s)$ and
$\beta(e)$ are the well-known QCD and QED $\beta$-functions.

The interaction vertices of the radion with the SM fields are
similar to those of the SM Higgs boson except for the anomaly
enhanced interactions with gluons and photons. This leads to the
corresponding relative enhancement in gluon and photon decay modes
and the gluon-gluon fusion production channel of the radion.
Various aspects of the decay and production properties of the
radion have been touched upon in a number of studies, including a
possible  mixing of the radion with the Higgs boson, which is
important for the radion and Higgs phenomenology
\cite{Giudice:2000av} - \cite{Gunion:2003px}.

After the discovery of the Higgs-like boson at the LHC
\cite{Higgs-discov} few studies have been curried out
investigating a possible impact of the radion as the second scalar
particle on the interpretation of the observed boson
\cite{Chacko:2012vm} - \cite{Jung:2014zga}. The details depend on
particular assumptions of the scenario under consideration, such
as which fields propagate in the bulk and which are localized on
the brane, what is the Higgs-radion mixing mechanism and what is
the mixing parameter etc. But generically the light radion with
mass below or above the observed 126 GeV boson is still not fully
excluded by all the EW precision constrains and the LHC data.

The fermion part of  Lagrangian (\ref{L1}) for on-shell fermions has
a very simple form
\begin{equation}
 L = - \sum_f \frac{r}{\Lambda_r} m_f \bar{f} f,
 \label{Lferm_onshell}
\end{equation}
being the same as for the Higgs boson with the replacement of the
scale $\Lambda_r$ by the Higgs vacuum expectation value,
$\Lambda_r \rightarrow v$. However, for the case of off-shell
fermions  Lagrangian (\ref{Lferm_onshell})  needs to be modified.
The corresponding Lagrangian can be written symbolically in the
following form (the explicit form of the Lagrangian is given in
Appendix 1) \cite{Boos:2007eg}
\begin{equation}
L = -\sum_f \frac{r}{\Lambda_r} [\frac{3i}{2}( (D_{\mu}\bar{f}) \gamma^{\mu}f - \bar{f}\gamma^{\mu} (D_{\mu} f)) + 4 m_f \bar{f} f] +...
 \label{Lferm_offshell}
\end{equation}
where $D_{\mu}$ are the SM covariant derivatives. Note that
Lagrangian (\ref{Lferm_offshell}) is non-trivial even for massless
fermions.

The aim of this short study is to investigate a possible influence
of these additional terms on the processes with tree level
diagrams involving the radion and off-shell fermion lines. The
extra terms in the Lagrangian lead to a momentum dependence in the
vertices potentially giving additional contributions if virtual
fermions participate in the process. For the main radion decay
processes the fermions are on-shell, and  these additional terms
in the vertices vanish, but this might not be the case for the
radion production.

However, it will be explicitly demonstrated that due to the gauge
invariance all the additional contributions to the amplitudes of
physical processes are canceled out, and the result for the matrix
element  is the same as  computed with only   Higgs-like
Lagrangian (\ref{Lferm_onshell}).
 One should stress that in order to get the mentioned cancellations
the trace $ T_{\mu}^{\mu}$ of the SM energy momentum tensor should
be computed including the SM covariant derivatives as given in
Appendix 1.  The corresponding Feynman rules containing new
4-point vertices are presented in Appendix 2.

We begin with a simple example of the radion production in the
radion-strahlung process. Then we give a more general explanation
demonstrating how the cancellation takes place and show a  few
more examples for the main radion  production processes at the LHC
and at a Linear Collider.

\section{The radion-strahlung as a simple example}
\label{sec:strahlung} 
\begin{figure*}[!h!]
\begin{center}
\includegraphics[height=3cm]{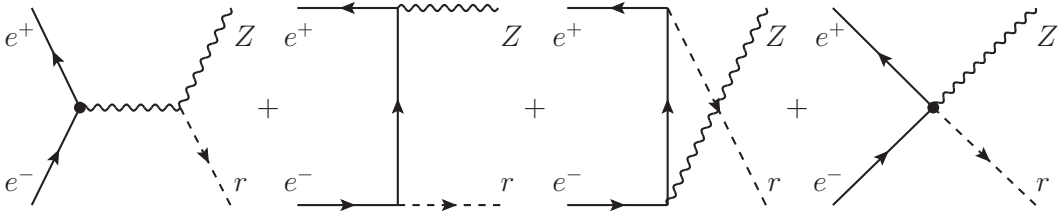}~~~
\end{center}
\caption[]{Feynman diagrams contributing to the radion production
in association with the Z boson in $e^+e^-$ collisions (the
radion-srtahlung process)} \label{fig:Radion-strahlung-diagrams}
\vspace*{-2mm}
\end{figure*}

We begin with a very simple example, the
radion production in association with $Z$ boson in $e^+e^-$
collisions called the radion-strahlung by analogy with the
well-known Higgs production process, the Higgs-strahlung. Even
neglecting the very small electron mass there are four
contributing Feynman diagrams at the lowest order shown in
Fig.\ref{fig:Radion-strahlung-diagrams}.

%
Note that the presence of the last 4-point diagram is a
consequence of the gauge invariance. The Feynman rules for the
vertices involved are given in  Appendix 2. Correspondingly the
diagrams give the following contributions to the process amplitude
\begin{equation}
M_1 = - 2i\, C\, \bar{e}^+(p_2)\,\Gamma_{\mu}\, e^-(p_1)\,\frac{1}{p^2 - M_Z^2}\, M_Z^2\, \varepsilon(p_Z)\, r(p_r)
\end{equation}
\begin{equation}
M_2 = - i\, C\,\bar{e}^+(p_2)\,[\frac{3}{2}\,(\cancel{k}+\cancel{p}_2)]\,\frac{\cancel{k}}{k^2}\,
\Gamma_{\mu}\,e^-(p_1)\,\varepsilon(p_Z)\,r(p_r)
\end{equation}
\begin{equation}
M_3 = - i\,C\,\bar{e}^+(p_2)\,\Gamma_{\mu}\,\frac{\cancel{q}}{q^2}\,[\frac{3}{2}\,(\cancel{q}-\cancel{p}_1)]
\,e^-(p_1)\,\varepsilon(p_Z)\,r(p_r)
\end{equation}
\begin{equation}
M_4 = + 3 i\, C\,\bar{e}^+(p_2)\,\Gamma_{\mu}\,e^-(p_1)\,\varepsilon(p_Z)\,r(p_r),
\end{equation}
where $C = \frac{1}{\Lambda_r}\frac{e}{2 \sin\theta_W \cos\theta_W}$, $p=p_1+p_2=p_r+p_Z$,
$ \Gamma_{\mu} = \gamma_{\mu} (2 \sin^2\theta_W - \frac{1-\gamma_5}{2})$, $k=p_2-p_r$, $q = p_r-p_1$.

Keeping in mind the Dirac equation $\cancel{p}_j \, e(p_j)\,= 0$ and the equality
$\cancel{q}\,\frac{\cancel{q}}{q^2}\,= 1$ it is easy to see that the sum  of two diagrams D2 and D3
is exactly canceled  out by the last diagram D4, so
\begin{equation}
M_2+M_3+M_4 = 0.
\end{equation}
Therefore the matrix element squared $|M|^2 = |M_1|^2$ for the
radion production is exactly equal to that for the Higgs boson,
and the cross section takes the well-known form with the
replacement $\Lambda_r \rightarrow v$ and $M_r \rightarrow M_h$
\begin{equation}
\sigma(e^+e^- \rightarrow r Z) =
\frac{M_Z^2}{\Lambda_r^2}\,
\frac{\alpha (8 \sin^4\theta_W - 4 \sin^2\theta_W +1}{24\,\sin^2\theta_W\,\cos^2\theta_W}\,
\frac{\sqrt{\lambda_r}}{4 s^2}\,
\frac{\lambda_r + 12 M_Z^2 s}{s - M_Z^2},
\end{equation}
where $\lambda_r = (M_Z^2 + M_r^2 -s)^2 - 4 M_Z^2 M_r^2$.

As one can see all the additional contributions are canceled out. The same property of the cancellation
of additional to the Higgs-like contributions takes place for the associated
radion and $W^{\pm}$ boson production, for example,
\begin{equation}
u \bar{d} \rightarrow r W^+
\end{equation}
where  there are also 4 contributing diagrams similar to those in
Fig.\ref{fig:Radion-strahlung-diagrams}.

\section{Cancellations of additional to the Higgs-like contributions in tree-level  amplitudes}
\label{sec:cancellations}
The observed  cancellation follows from
the structure of the fermion current with the emission of the
radion and a number of gauge bosons.

\begin{figure*}[!h!]
\begin{center}
\includegraphics[height=2.5cm]{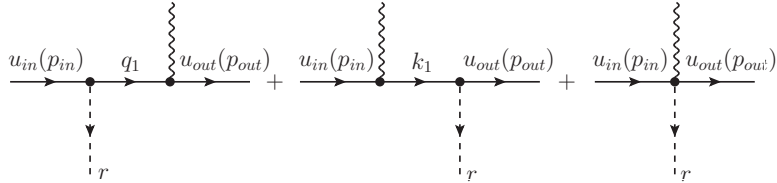}
\end{center}
\caption[]{Fermion current radiating the radion and an SM vector
gauge boson} \label{fig:Radion-1V-diagrams} \vspace*{-2mm}
\end{figure*}

Let us begin with a simple case with just one gauge boson emission
as shown in  Fig.\ref{fig:Radion-1V-diagrams}. It corresponds to
the radion-strahlung process considered in the previous section
with Z boson emission. However the cancellation property is valid
for any emitted SM gauge boson and for any massive SM fermion.

In order to see that it is instructive to rewrite the fermion-radion vertex in the following way
$$\frac{i}{\Lambda_r}\,[\frac{3}{2}\,(\cancel{p}_{out}\, + \,\cancel{p}_{in}) - 4\,m_f] =$$
$$\frac{i}{\Lambda_r}\,[\frac{3}{2}\,(\cancel{p}_{out}\, - \, m_f)\,
                  + \,\frac{3}{2}\,(\cancel{p}_{in}\, - \, m_f) - m_f] =$$
\begin{equation}
\frac{i}{\Lambda_r}\,[\frac{3}{2}\,S^{-1}(p_{in})\, + \,\frac{3}{2}\,S^{-1}(p_{out}) - m_f],
\label{vertex-f-r}
\end{equation}
where $S^{-1}(p)$ is the inverse function to the propagator
\footnote{All the relevant constants from propagators and vertices
are included into the common factors of the amplitudes.}
$$S(p) = \frac{\cancel{p}\,+\,m_f}{p^2 \,-\,m_f^2}. $$
Note that the last term proportional to $m_f$ in vertex
(\ref{vertex-f-r}) is exactly the same as for the Higgs boson with
the obvious replacement $\Lambda_r \to v$.

The diagrams in Fig.\ref{fig:Radion-1V-diagrams} can be expressed
as follows
\begin{equation}
D_1 = -i\, C\, \bar{u}_{out}(p_{out}) \, \Gamma_{\mu} \, S(k_1)\,
[\frac{3}{2}\,(S^{-1}(k_1)\, + \,S^{-1}(p_{in})) - m_f]\,  u_{in}(p_{in})
\end{equation}
\begin{equation}
D_2 = -i\,C\, \bar{u}_{out}(p_{out}) \,
[\frac{3}{2}\,(S^{-1}(p_{out})\, + \,S^{-1}(q_1)) - m_f]\, S(q_1)\,\Gamma_{\mu} \, u_{in}(p_{in})
\end{equation}
\begin{equation}
D_3 = +i\,3\,C\,\bar{u}_{out}(p_{out})\,\Gamma_{\mu} \, u_{in}(p_{in})
\end{equation}
where the constant C includes the  product of all the factors of
the vertices involved, $\Gamma_{\mu}$ is the Lorentz part of the
fermion-boson and fermion-boson-radion vertices, which is the same
in two vertices coming from the same covariant derivative. The
initial and the final spinors correspond to the same fermion for
the neutral fermion current and they are different for the charged
current. Once more one can easily see that using the Dirac equation for
$in$ and $out$ fermion states and summing up
$D_1+D_2+D_3$  the only remaining contribution is proportional to
the fermion mass and is  the same as that for the Higgs boson.

\begin{figure*}[!h!]
\begin{center}
\includegraphics[height=2.5cm]{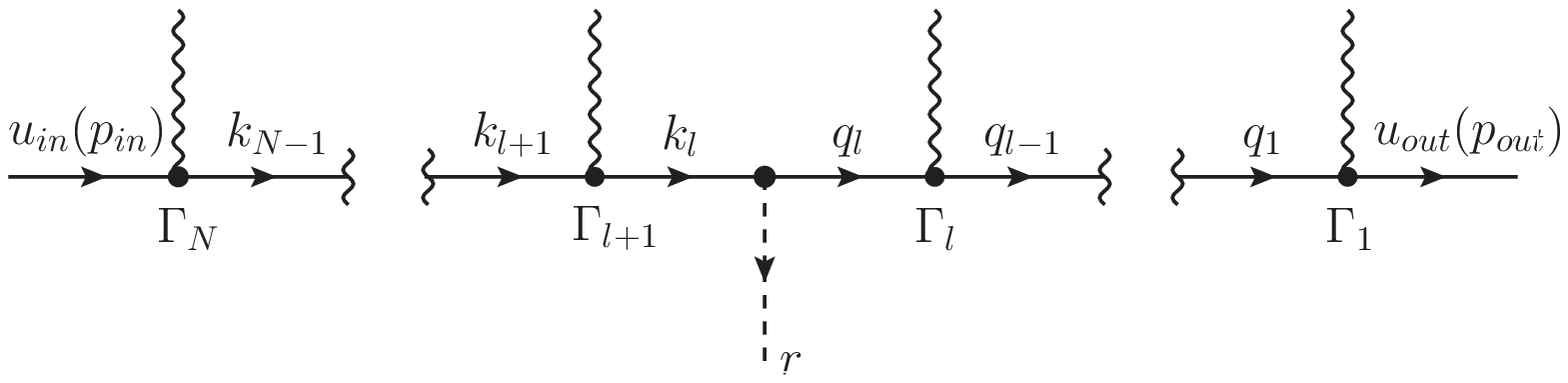}\\
\vspace{3mm}
\includegraphics[height=2.5cm]{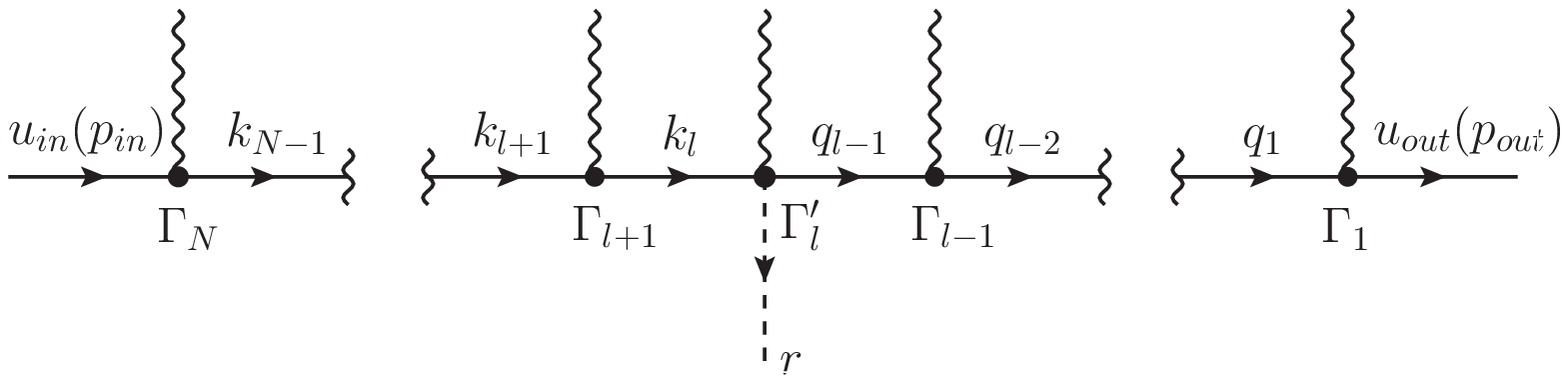}
\vspace*{-3mm}
\end{center}
\caption[]{Fermion current radiating the radion and $N$ SM vector
gauge bosons in a three-point vertex (upper graph) and in a
four-point vertex (lower graph) corresponding to the contributions
$M^l$ and $M^\prime_l$ respectively.}
\label{fig:Radion-NV-diagrams} \vspace*{-2mm}
\end{figure*}

This property of the cancellation of all the contributions
additional to the Higgs-like part can be generalized to an
arbitrary number, say $N$,  of  gauge bosons emitted from the
fermion line in association with the radion as shown in
Fig.\ref{fig:Radion-NV-diagrams}.

The sum of all diagrams can be split into parts containing the
fermion-fermion-radion vertex and the boson-fermion-fermion-radion
vertex
\begin{equation}
M_{N vector\, bosons} \, = \, M_0\, +\, \sum_{l=1}^{N} (M_l\, +
\,M^\prime_l),
\end{equation}
where
\begin{equation}
\begin{split}
M_l \sim \, i^{2N+1} \bar{f}_{out}(p_{out})\,[\prod_{j=1}^l \,
\Gamma_{\mu_j}^j S(q_{j})]
[-\frac{3}{2}\,(\,S^{-1}(q_l)\, + \,S^{-1}(k_l)\,) + m_{f_l}]&   \\
[\prod_{j=1+1}^N \, S(k_{j-1}) \,\Gamma_{\mu_j}^j]\,
f_{in}(p_{in}),&
\end{split}
\label{Ml}
\end{equation}
 for $l=1$,...,$N-1$
\begin{equation}
M_0 \sim  \, i^{2N+1} \bar{f}_{out}(p_{out})\,
[-\frac{3}{2}\,(\,S^{-1}(p_{out})\, + \,S^{-1}(k_{0})\,) + m_{f_{out}}]\,
[\prod_{j=1}^N \,  S(k_{j-1})\Gamma_{\mu_j}^j] f_{in}(p_{in}),
\label{Ml0}
\end{equation}
\begin{equation}
M_N \sim  \, i^{2N+1} \bar{f}_{out}(p_{out})\,[\prod_{j=1}^N \, \Gamma_{\mu_j}^j S(q_j)]
[-\frac{3}{2}\,(\,S^{-1}(q_N)\, + \,S^{-1}(p_{in})\,) + m_{f_{in}}]
f_{in}(p_{in}),
\label{MlN}
\end{equation}
\begin{equation}
M^\prime_l \sim  \, i^{2N-1} \bar{f}_{out}(p_{out})\,[\prod_{j=1}^{l-1} \, \Gamma_{\mu_j}^j S(q_{j})]
[-3\,\Gamma_{\mu_l}^l]\,[\prod_{j=l+1}^{N} \, S(k_{j-1})\Gamma_{\mu_j}^j] f_{in}(p_{in}),
\label{Mprime-l}
\end{equation}
for  $l= 2$,...,$N-1$
\begin{equation}
M^\prime_1 \sim  \, i^{2N-1} \bar{f}_{out}(p_{out})\,[-3\,\Gamma_{\mu_1}^1]
[\prod_{j=2}^N \, S(k_{j-1})\Gamma_{\mu_j}^j] f_{in}(p_{in})
\label{Mprime-l1}
\end{equation}
\begin{equation}
M^\prime_N \sim  \, i^{2N-1} \bar{f}_{out}(p_{out})\,
[\prod_{j=1}^{N-1} \, \Gamma_{\mu_j}^j S(q_j)]\,[-3\,\Gamma_{\mu_N}^N] f_{in}(p_{in}),
\label{Mprime-lN}
\end{equation}
$\Gamma_{\mu_j}^j$ is the SM fermion and gauge boson vertex, $k_l
= p_{in} - \sum_{j=l}^N p_j^V$, $q_0 = p_{in}$, $q_l=k_l-p_r =
p_{out} + \sum_{j=1}^l p_j^V$, $p_j^V$ is the vector boson
momentum, the upper index $V$ standing for any SM gauge vector
boson (Z, W, photon, gluon), $p_r$ is the radion momentum, $m_{f_l}$,
$m_{f_{in}}$, $m_{f_{out}}$ are the masses of the internal,
initial and final state fermions which may be different in the
case of the charged current. Note that the factors  $i^{2N+1}$ and
$i^{2N-1}$ are different due to the difference by one of the
number of the propagators and vertices involved into expressions
(\ref{Ml}), (\ref{Ml0}), (\ref{MlN}) and
(\ref{Mprime-l}), (\ref{Mprime-l1}), (\ref{Mprime-lN}). This leads
to a relative ($-1$) sign between the two contributions.

From  formulas (\ref{Ml})--(\ref{Mprime-lN})  one can
write simple relations taking into account the relation
$S^{-1}(q_j)\times S(q_j) =1$ and the Dirac equations for $in$ and
$out$ fermions
\begin{equation}
M_l = M_l^H - \frac{1}{2} M^\prime_l - \frac{1}{2} M^\prime_{l+1}
\label{relation1}
\end{equation}
\begin{equation}
M_0 = M_0^H - \frac{1}{2} M^\prime_1
\label{relation2}
\end{equation}
\begin{equation}
M_N = M_N^H - \frac{1}{2} M^\prime_N,
\label{relation3}
\end{equation}
where the Higgs-like contributions $M_0^H$, $M_N^H$, $M_N^H$ labeled with the $H$ symbol
being proportional to the fermion masses have the form which follows from (\ref{Ml}),(\ref{Ml0}),(\ref{MlN})
\begin{equation}
\begin{split}
M_l^H \sim & \, i^{2N+1} \bar{f}_{out}(p_{out})\,[\prod_{j=1}^l \,  \Gamma_{\mu_j}^j S(q_{j})]
[ m_{f_l}] [\prod_{j=1+1}^N \, S(k_{j-1}) \,\Gamma_{\mu_j}^j]\,  f_{in}(p_{in}),
\end{split}
\label{MlH}
\end{equation}
\begin{equation}
M_0^H \sim  \, i^{2N+1} \bar{f}_{out}(p_{out})\,
[m_{f_{out}}]\,
[\prod_{j=1}^N \,  S(k_{j-1})\Gamma_{\mu_j}^j] f_{in}(p_{in}),
\label{Ml0H}
\end{equation}
\begin{equation}
M_N^H \sim  \, i^{2N+1} \bar{f}_{out}(p_{out})\,[\prod_{j=1}^N \, \Gamma_{\mu_j}^j S(q_j)]
[ m_{f_{in}}]
f_{in}(p_{in}),
\label{MlNY}
\end{equation}
Summing up the left and the right hand sides of  equation
(\ref{relation1}) from $1$ to $N-1$ and adding the left and the
right hand sides of  equations (\ref{relation2}) and
(\ref{relation3}) one gets the following equality
\begin{equation}
\sum_{l=0}^N M_l + \sum_{l=1}^N M^\prime_1 = \sum_{l=0}^N M_l^H
\label{main}
\end{equation}
The result in formula (\ref{main}) means that the sum of all the
contributions leads to only the Higgs-like type of the
contribution and all the other parts are canceled out explicitly.

\section{Generalization to the loop case}
\label{sec:loop-case} The main result of the previous section can
be easily generalized to the loop case. First of all, one should
mention that in the  way of proving  equality (\ref{main}) nothing
special was required for boson lines in the vertices. The boson
lines in  Fig.\ref{fig:Radion-NV-diagrams} could correspond to
real particles or virtual propagators in the loops. In this sense
all the above considerations are valid for both real and virtual
 gauge bosons emitted from the fermion current.

In the case of a fermion loop one can follow a similar logic as
was given above. In the loop case there is an additional fermion
propagator instead of the external spinors at the tree level.
Correspondingly the expressions for various fermion loop
contributions with emission of $N$ gauge bosons have the following
form by analogy with (\ref{Ml}),(\ref{MlN}) and
(\ref{Mprime-l}),(\ref{Mprime-l1}),(\ref{Mprime-lN}).
\begin{equation}
\begin{split}
M_l \sim  \, i^{2N+1} Tr \left\{ \,[\prod_{j=1}^l \,  \Gamma_{\mu_j}^j S(q_{j})]
[-\frac{3}{2}\,(\,S^{-1}(q_l)\, + \,S^{-1}(k_l)\,) + m_{f_l}] \right. \\
\left. [\prod_{j=1+1}^N \, S(k_{j-1}) \,\Gamma_{\mu_j}^j]\,S(p) \right\},
\end{split}
\label{Ml-loop}
\end{equation}
for $l=1$,...,$N-1$,
\begin{equation}
M_N \sim  \, i^{2N+1}Tr \left\{\, [\prod_{j=1}^N \, \Gamma_{\mu_j}^j S(q_j)]
[-\frac{3}{2}\,(\,S^{-1}(q_N)\, + \,S^{-1}(p_{in})\,) + m]\,S(p) \right\},
\label{MlN-loop}
\end{equation}
\begin{equation}
M^\prime_l \sim  \, i^{2N-1}Tr \left\{ \,[\prod_{j=1}^{l-1} \, \Gamma_{\mu_j}^j S(q_{j})]
[-3\,\Gamma_{\mu_l}^l]\,[\prod_{j=l+1}^{N} \, S(k_{j-1})\Gamma_{\mu_j}^j]\,S(p) \right\},
\label{Mprime-l-loop}
\end{equation}
for  $l= 2$,...,$N-1$
\begin{equation}
M^\prime_1 \sim  \, i^{2N-1} Tr \left\{\,[-3\,\Gamma_{\mu_l}^l]
[\prod_{j=2}^N \, S(k_{j-1})\Gamma_{\mu_j}^j] \,S(p) \right\},
\label{Mprime-l1-loop}
\end{equation}
\begin{equation}
M^\prime_N \sim  \, i^{2N-1} Tr \left\{\,
[\prod_{j=1}^{N-1} \, \Gamma_{\mu_j}^j S(q_j)]\,[-3\,\Gamma_{\mu_N}^N] \,S(p) \right\}.
\label{Mprime-lN-loop}
\end{equation}
where the momenta are expressed as $k_l = p - \sum_{j=l}^N p_j^V$,
$q_l=k_l-p_r = p + \sum_{j=1}^{l-1} p_j^V$. Note that the
contribution $M_0$ is not present now since it coincides with
$M_N$.

In the same manner as in the previous section we get
\begin{equation}
M_l = M_l^H - \frac{1}{2} M^\prime_l - \frac{1}{2} M^\prime_{l+1}
\label{relation1-loop}
\end{equation}
\begin{equation}
M_N = M_N^H - \frac{1}{2} M^\prime_N - \frac{1}{2} M^\prime_{1},
\label{relation3-loop}
\end{equation}
From  relations (\ref{relation1-loop}) and (\ref{relation3-loop})
it is easy to show that
\begin{equation}
\sum_{l=1}^N M_l = \sum_{l=1}^{N-1} + M_N = \sum_{l=1}^N M_l^H - \sum_{l=1}^N M^\prime_1
\end{equation}
and therefore one gets once again the equality
\begin{equation}
\sum_{l=1}^N M_l + \sum_{l=1}^N M^\prime_1 = \sum_{l=1}^N M_l^H ,
\label{main-loop}
\end{equation}
which demonstrates that all the contributions except for the
Higgs-like type are canceled out in the case of a fermion loop.

\section{Conclusions}
\label{sec:conclusions} In all the main production processes of
the radion, the lowest KK mode of a combination of the spin-0
metric fluctuation and  the scalar stabilizing field, there are
contributing Feynman diagrams involving off-shell fermions. The
radion is emitted from various  fermion currents containing
off-shell fermion propagators in the radion-strahlung or the
vector boson fusion production in $e^+e^-$ collisions as well as
in the vector boson fusion, associated radion and vector boson
production, associated radion and the top-quark pair production
processes in hadron collisions at the LHC. The off-shell fermions
participate in fermion loops for gluon-gluon fusion production
process at the LHC and in $gg$, $\gamma\gamma$, $\gamma Z$ decay
modes of the radion.
 The additional to the Higgs boson case non-trivial vertices (even for massless
fermions)  of fermion-radion interactions follow from the
structure of the trace of the gauge invariant SM energy-momentum
tensor as given in the Appendices~1,2.

We have shown that all the contributions to perturbative
amplitudes of physical processes due to corresponding additional
terms are canceled out for both massless and massive off-shell
fermions. We demonstrated, first, the cancellation in a simple
example of the radion-strahlung process in $e^+e^-$ collisions.
Then we presented a general proof of the cancellation for an
arbitrary fermion current radiating the radion and any number of
the SM gauge bosons. This proof was generalized to the case of the
amplitudes containing closed fermion loops and an arbitrary number
of the gauge bosons.

The proof also means that the terms with the covariant derivatives
of the fermion fields in Lagrangian (\ref{Lferm_offshell}) can be
replaced by the mass terms in accordance with the equation of
motion for the fermion fields, which is not {\it a priori}
justified in the gauge theory, where the quantization is performed
with the help of the path integration.

Thus the additional fermion-radion terms in the interaction Lagrangian do
not alter any production and decay properties of the radion
compared to those of the Higgs boson.

One should mention that the observed cancellation property, being
a result of the gauge invariant structure of the SM
energy-momentum trace, is also valid for any scalar particle (not
only the radion) which interacts with the SM particles via the
trace.

\section{Acknowledgments}
The work was supported by   grant 14-12-00363 of Russian Science
Foundation. The authors are grateful to  M.~Smolyakov and
I.~Volobuev for useful discussions and critical remarks.

\section{Appendix 1}
\label{sec:trace}
\textbf{Trace of the SM energy-momentum tensor.}\\
The trace of the SM energy-momentum tensor calculated as the
variation  of the SM Lagrangian with respect to the metric
\cite{Landau,BD} can be written in the unitary gauge as follows:
\begin{eqnarray*}
T_{\mu }^{\mu } & =& -\left(\partial _{\mu }
h\right)\left(\partial ^{\mu } h\right)+2m_{h}^{2} h^{2}
\left(1+\frac{h}{2v} \right)^{2}
- 2m_{W}^{2} W_{\mu }^{+} W^{\mu \, -} \left(1+\frac{h}{v}\right)^{2} -m_{Z}^{2} Z_{\mu } Z^{\mu } \left(1+\frac{h}{v} \right)^{2} \\
&+& \sum _{f}\left\{-\frac{i3}{2} \left[\bar{f}\gamma ^{\mu }
\left(\partial _{\mu } f\right)-\left(\partial _{\mu }
\bar{f}\right)\gamma ^{\mu } f\right]+4m_{f} \bar{f}f\right\} +
\frac{4h}{v} \sum _{f}m_{f} \bar{f}f-3eA_{\mu } \sum _{f}q_{f}
\bar{f}\gamma ^{\mu } f  \\
&-&  \frac{3}{2}\frac{m_{Z}}{v} Z_{\mu } \sum _{f}\bar{f}\gamma ^{\mu } \left[a_{f} +b_{f} \gamma _{5} \right]f
- \frac{3}{\sqrt{2}}\frac{m_{W}}{v} \left(W_{\mu }^{-} \bar{\nu }_{j} U_{jk}^{PMNS} \gamma ^{\mu } \left[1-\gamma _{5} \right]e_{k} +h.c. \right) \\
&-& \frac{3}{\sqrt{2}}\frac{m_{W}}{v}  \left(W_{\mu }^{-} \bar{u}_{j} \gamma ^{\mu } \left[1-\gamma _{5} \right]V_{jk}^{CKM} d_{k} +h.c. \right)
- 3g_{c} \left(\bar{u}_{j} \gamma ^{\mu } \hat{G}_{\mu } u_{j} +\bar{d}_{j} \gamma ^{\mu } \hat{G}_{\mu } d_{j} \right) \\
\end{eqnarray*}
\begin{eqnarray*}
&+& \frac{\beta(e)}{2e } F_{\mu \nu } F^{\mu \nu
}+\frac{\beta(g_{s})}{2g_{s} } G_{\mu \nu }^{a b } G_{a b }^{\mu
\nu },\phantom{aaaaaaaaaaaaaaaaaaaaaaaaaaaaaaaaaaaaaaa}
\end{eqnarray*}
where $U^{PMNS}$ is the
Pontecorvo-Maki-Nakagawa-Sakata matrix, $V^{CKM}$ is the
Cabibbo-Kobayashi-Maskawa matrix. The last two terms take the
anomalies into account, they describe the interaction of the
radion with the photon and the gluon fields.

\section{Appendix 2}
\label{sec:Frules}

\begin{tabbing}
\textbf{Interaction vertices of radion and SM fields}\\
\\
Radion and two fermions:\\
\\
$r(k_{r})\bar{f}(p_{\bar{f}} )f(p_{f} )$\qquad\qquad\qquad  $\frac{i}{\Lambda _{r} } \left\{\frac{3}{2}
\left[\rlap{$/$}p_{\bar{f}} -\rlap{$/$}p_{f} \right]+4m_{f} \right\}=\frac{i}{\Lambda _{r} }
\left\{\frac{3}{2} \left[\left(\rlap{$/$}p_{\bar{f}} +m\right)-\left(\rlap{$/$}p_{f} -m\right)\right]+m_{f} \right\}$\\
\\
Radion, two fermions and photon:\\
\\
$r(k_{r} )\gamma (k_{\gamma } )\bar{f}(p_{\bar{f}} )f(p_{f} )$\qquad\qquad\qquad\qquad  \=  $-i\frac{3eq_{f} }{\Lambda _{r } } \gamma
^{\mu } $\\
\\
Radion, two fermions and $Z$-boson:\\
\\
$r(k_{r} )Z(k_{Z} )\bar{f}(p_{\bar{f}} )f(p_{f} )$\>  $-i\frac{1}{\Lambda _{r } } \frac{3m_{Z}}{2v} \gamma ^{\mu } \left[a_{f} +b_{f} \gamma _{5} \right]$\\
\\
Radion, two fermions and $W$-boson:\\
\\
$r(k_{r} )W(k_{W} )\bar{u}(p_{u} )d(p_{d} )$\>  $-i\frac{1}{\Lambda _{r } } \frac{3m_{W}}{\sqrt{2}v} V_{jk}^{CKM} \gamma ^{\mu } \left[1-\gamma _{5} \right]$\\\\
$r(k_{r} )W(k_{W} )\bar{\nu }(p_{\nu } )e(p_{e} )$\> $ -i\frac{1}{\Lambda _{r } } \frac{3m_{W}}{\sqrt{2}v } U_{jk}^{PMNS} \gamma ^{\mu } \left[1-\gamma _{5} \right]$\\
\\
Radion, two fermions and Higgs-boson:\\
\\
$r(k_{r} )h(k_{h} )\bar{f}(p_{\bar{f}} )f(p_{f} )$\>  $i\frac{1}{\Lambda _{r } } \frac{4m_{f} }{v } $\\
\\
Radion and two Higgs-bosons:\\
\\
$r(k_{r} )h(p_{1} )h(p_{2} )$\>  $ i\frac{1}{\Lambda _{r } } \left\{p_{1\mu } p_{2} ^{\mu } -2m_{h}^{2}
\right\}$\\
\\
Radion and three Higgs-bosons:\\
\\
$r(k_{r} )h(p_{1} )h(p_{2} )h(p_{3} )$\>  $-i\frac{1}{\Lambda _{r } } \frac{12m_{h}^{2} }{v } $\\
\\
Radion and four Higgs-bosons:\\
\\
$r(k_{r} )h(p_{1} )h(p_{2} )h(p_{3} )h(p_{4} )$\>  $-i\frac{1}{\Lambda _{r } } \frac{12m_{h}^{2} }{v^{2} } $\\
\\
Radion and two $Z$-bosons:\\
\\
$r(k_{r} )Z(p_{1} )Z(p_{2} )$\> $  -i\frac{2m_{Z}^{2} }{\Lambda _{r } } g^{\mu \nu } $\\
\\
Radion, two $Z$-bosons and Higgs-boson:\\
\\
$r(k_{r} )Z(p_{1} )Z(p_{2} )h(p_{3} )$\>  $-i\frac{1}{\Lambda _{r } } \frac{4m_{Z}^{2} }{v } g^{\mu \nu } $\\
\\
Radion, two $Z$-bosons and two Higgs-bosons:\\
\\
$r(k_{r} )Z(p_{1} )Z(p_{2} )h(p_{3} )h(p_{4} )$\> $ -i\frac{1}{\Lambda _{r } } \frac{4m_{Z}^{2} }{v^{2} } g^{\mu \nu } $\\
\\
Radion and two $W$-bosons:\\
\\
$r(k_{r} )W(p_{1} )W(p_{2} )$\>   $-i\frac{2m_{W}^{2} }{\Lambda _{r } } g^{\mu \nu } $\\
\\
Radion, two $W$-bosons and Higgs-boson:\\
\\
$r(k_{r} )W(p_{1} )W(p_{2} )h(p_{3} )$\>  $-i\frac{1}{\Lambda _{r } } \frac{4m_{W}^{2} }{v }
g^{\mu \nu } $\\
\\
Radion, two $W$-bosons and two Higgs-bosons:\\
\\
$r(k_{r} )W(p_{1} )W(p_{2} )h(p_{3} )h(p_{4} )$\> $-i\frac{1}{\Lambda _{r } } \frac{4m_{W}^{2} }{v^{2} } g^{\mu \nu } $\\
\\

\end{tabbing}


\bibliographystyle{model1-num-names}



\end{document}